# A Fair Method for Distributing Collective Assets in the Stellar Blockchain Financial Network

**Kiarash Shamsi[1], Mohammad Javad Shayegan [2]**
[1]Department of Computer Engineering, University of Science and Culture, Tehran, Iran
[2]Department of Computer Engineering, University of Science and Culture, Tehran, Iran

Corresponding author: Mohammad Javad Shayeganfard (e-mail: shayegan@usc.ac.ir).

**ABSTRACT** The financial industry is a pioneer in Blockchain technology. One of the most popular platforms in Token-based banking is the flexible Stellar platform. This platform is open-source, and today, its wide range of features makes it possible for many countries and companies to use it in cryptocurrency and Token-based modern banking. This network charges a fee for each transaction. As well, a percentage of the net amount is generated as the inflation rate of the network due to the increased number of tokens. These fees and inflationary amounts are aggregated into a general account and ultimately distributed among members of the network on a collective vote basis. In this mechanism, network users select an account as the destination for which they wish to transfer assets using their user interface, which is generally a wallet. This account could be the account of charities that need this help. It is then determined the target distribution network based on the voting results of all members. One of the challenges in this network is the purposeful and fair distribution of these funds between accounts. In this paper, the first step is a complete infrastructure of a Stellar financial network that will consist of three network-based segments of the core network, off-chain server, and wallet interface. In the second step, a context-aware recommendation system will be explored and implemented as a solution for the purposeful management of payroll account selection. The results of this study concerning the importance of the purposeful division of collective assets and showing a context-aware recommendation system as a solution to improve the process of stellar users' participation in the voting process by effectively helping them in choosing an eligible destination

**INDEX TERMS** Blockchain, Cryptocurrency, Token-Based Banking, Stellar, Fair Distribution

## I. INTRODUCTION

After introducing Blockchain, different applications and industries have started to use different types of networks based on their needs. The payment industry has shown a greater willingness to use this technology in its new systems because of its proximity to the nature of it. The financial industry's 60% share of the total investment in Blockchain technology is evident [1]. The financial industry is using this technology in areas such as customer identification, asset management, asset tokenizing, international payments, and many more.

Benefits such as transparency, reliability, and consistency of information in Blockchain-based networks allow financial and credit institutions to offer a variety of token assets and provide diverse products to their customers [2]. Although financial exchange in a completely open space offers many benefits, due to the nature of banking and the competitive environment between banks, these institutions tend to monitor the existing network and also evaluate the behavior of users [3]. One of the most popular Blockchain-based financial networks is the Stellar Financial Network. As the network is free and open-source, using it to develop new systems is very easy and cost-effective. Also, a complete platform consisting of many modern banking features required by the payment industry is considered in Stellar. These features have made Stellar a growing system in various countries for use in the core of Blockchain-based financial networks [4]. One of the most essential and unique features of this network is the presentation of a model of aggregation among members of the network as part of a voting process for the division of collective assets, including annual inflation percentages, commission fees, and charitable donations. In this process, each user selects an address as the destination for the money sharing and announces to the network. Then, the network divides the amounts collected on the collective account based on the basic cryptocurrency among those who received more collective votes from users [5]. Most banks, on the other hand, have charity accounts that users can choose to share as commissions to help them collectively. In most cases, users find it challenging to select the entity they want to participate in the voting process and usually choose the most popular ones. This issue creates an unfair division of assets

## II. RESEARCH BACKGROUND

| | | | |
|---|---|---|---|
| | Table 1.Comparison of previous works | | |
| [6] | Assume all members equally | 2018 | Machine learning prediction system |
| [7] | The components are not treated equally, and a fair threshold is defined as p for each segment | 2015 | Classification system |
| [8] | In addition to the p% method, it indicates the level of competence of each member in their own department | 2016 | Information Processing System |
| [9] | Scoring is entirely independent based on a number of its features | 2018 | Information Processing System |
| [10] | And divides the selection priorities equally | 2012 | operating system |
| [11] | Suggestions ranking based on their performance and priority | 2017 | Information and knowledge management system |

between different needy entities. Such decisions are usually made based on a series of sensitive features such as the degree of the reputation of the institution, the field of work, the dependencies of the institution, etc., which lead to the unfair elimination of the institutions that are needy in the collective voting process.

Because of the importance of a fair and targeted distribution of these assets, which comprise very high amounts in a national financial system, it is necessary to provide a strategy to help users for fairer voting. In this paper, the design and implementation of a complete financial system based on the Stellar Blockchain Network are presented, a solution using a context-aware recommender system to address the challenge of collective asset sharing. The system consists of three general parts that form it together.

1. The user interface, which mediates the user's relationship with the system, as well as the recording of user transactions. This part is also known as the wallet.
2. The core system of computing and generating suggestions to the user to connect to the user interface
3. The Stellar-based Blockchain network, which is the central core of network transaction logs. Together, they form the whole integrated system. Using a user interface connected to the Stellar Blockchain network, which is the same user wallet, the system provides a section for users to participate in this process. This section provides targeted suggestions for optimal selection in the process using an off-chain text-based recommender system that includes dimensions of the network and the user. As well, this system without having to know the complex public addresses enables the user to contribute easily to the cumulative voting process.

Collective fair voting in the Stellar network is still in its infancy, and so far, no solution has been developed to improve the fairness of the selection of division targets. However, generally, different definitions of fair algorithms have been proposed that are usable in the recommender systems used in this research, and it will be the basis of our work.

In [6], they propose a solution called awareness. This solution states that to select members in a system, we do not consider their sensitive and unique features as a whole, assuming all members are the [6]. In other words, if $A \in \{0, 1\}$ is a sensitive characteristic of, for example, the number of people employed by a charity, 1 represents more than 100 and 0 represents less than 100, and $C := c(X, A) \in \{0,1\}$ as the selection function of this institution in the recommender system, according to this strategy. Formula 1 shows the selection function of importance equality.

| |
|---|
| C=c (x, A) = c(X) |

**Formula 1.** The selection function of importance equality

This means that the importance of large and small enterprises is the same, and these characteristics have no influence on their selection.

[7] presents a solution called the p% law. In this rule, if $A \in \{0, 1\}$ is the same characteristic of the persons served by the institutions, the institutions with $A = 1$ are large corporations, and the institutions with $A = 0$ are small corporations. In this way, these entities are not treated equally, and a threshold is defined as p, which specifies the threshold for fair selection. This limit is usually divided by 80%, which means selecting from group A =

1 in 80% of choices and choosing from group A = 0 in 20% of choices. This rule is known as Four-Fifth Law [7]. The disadvantage of this method when using the fixed four-fifths rule is that although the choice between the two groups seems to be somewhat fair, there is no criterion for fair selection among the category itself so that if each time categories are selected randomly according to this algorithm, the result is fair.

In [8], a solution is presented, which is called Equality of Opportunities. In this solution, in addition to the preceding sections, a set is defined as Y ∈ {0, 1}. This set can be non-binary and contains many variables and conditions. This collection reflects the level of competence of each member in their department. In other words, in each of the sets discussed with A = 0 or A = 1, some members are more important to being elected. For example, they meet the needs of patients that helping them is more important [8]. In this solution, the goal is to provide equal opportunity to qualified members in two different areas. If P0 is the probability of being selected from the group with A = 0 and P1 is the probability of being selected from the group with A = 1, this solution is shown in formula 2:

$$P_0[C = 1 | Y = 1] = P_1[C = 1 | Y = 1]$$

**Formula 2.** Opportunity equality strategy

In [9], inspired by (Cynthia Dwork et al., 2012), a solution is proposed which is called individual justice, which means for each member, regardless of the group in which he belongs,

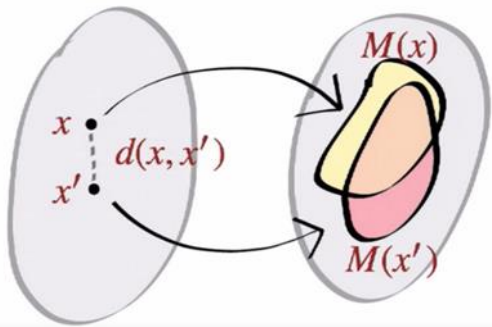

**Figure 1.** x and x2 are mapped to two different selection spaces by the function (M)

and independently based on a number of its properties called
input property vectors, as well as using a function called the metric function, the degree of selectivity of this member will be determined. In this solution, the metric function design according to the problem conditions and inputs, will be the most important part of the solution. Figure 1 shows the x and x2 feature vector mapped to two different selection spaces by the function (M) [9].

These methods can lead to a fair process for ranking available suggestions to the users in a recommender system. These systems calculate points for ranking and sorting the list of suggestions for the user by calculating several attributes by type and position. User feedback to this sorted list can help improve the results of the recommender system. Of course, there are also traditional approaches to create justice that is used in operating systems to schedule processes. The simplest of these is the Round-Robin method. This scheduling method has limitations. Including that many of the condition-based requirements are not concerned to provide an optimal suggestion, and selection priorities are shared equally [10]. As a result, using such algorithms for use in fair recommender systems will not yield excellent efficiency.

In [11], a method is presented to rank suggestions based on their efficiency and priority using some existing factors, relevant to conditions. In this system, any option obtains a score called performance score by using a set of features and factors. This score is constantly updated, and the user is offered the appropriate suggestion [11]. This system somehow creates justice by using equality of opportunity. Table 1 presents an overview of the different methods and contexts of each for use in this study.

Although a lot of works have been done in various fields in the fair algorithms field, there has been no work in the field of collective division Blockchain algorithms in the Blockchain networks in particular Stellar. Although earlier articles have been studied on recommender systems concerning the nature of Blockchain-based networks in the financial field and the need for improving the purposeful division of collective assets by using the concept of voting, there is not given any approach in this area yet. In voting-based systems, purposeful participation of all members and helping them to choose the best are the goals of the system, and due to the nature of voting, one can focus only on the presenting suggestion to the users, and their final choice cannot be changed.

### III. RESEARCH METHOD

The overall process of this research, which includes implementing and connecting a destination selection improvement system in our stellar financial network, as a solution to the distributing collective assets challenge is illustrated in figure 2:

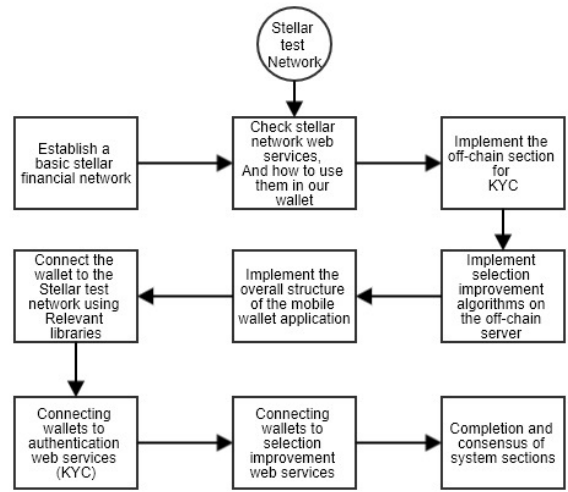

**Figure 2.** overall Research process

In general, the destination selection improvement system based on the Stellar Blockchain Network consists of three critical sections. These sections include:

1- Intrachain network based on the Stellar Blockchain system that provides the entire network base structure for use in the user interface, which is the basic token wallet. This structure includes the basic protocols and libraries that the Stellar network has provided to developers for open source use.

2. An off-chain process that is used to collect user identification information, basic information record for use in the selection improvement system, as well as the platform of the performance of the collective distribution destination selection system. The algorithms needed to use the selection improvement system are also implemented in this section.

3. Interface connectivity and integration of the above sections to use and deliver results to the user which in this research is an Android-based mobile wallet. This wallet is a link between the Blockchain network and the off-chain system that is used to establish an end-user relationship with the system and plays a vital role in data collection and overall system development. In the following, we give a detailed overview of each of the above. Figure 3 shows the system overview.

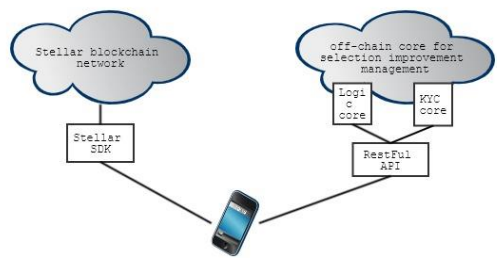

**Figure 3.** Schematic of the system of collective distribution improvement destination selection

A. *Blockchain Chain Transaction Service Provider*

Building a Stellar-based network that is capable of fully delivering all the services in the Stellar system requires a considerable amount of time and money to provide and configure many servers. This makes it impossible for independent developers or researchers to be able to provide these features to use and study this network. Therefore, to solve this problem, Stellar has developed a very practical solution. Stellar provides a network containing 3 nodes that form a complete Blockchain network. This network is generally separated from the original Stellar network on which the Lumen cryptocurrency is located. It has also developed a web application for the initial use of network APIs under HTTP protocol. This app was created to use and test many APIs on the network. Stellar has named this network the Stellar Test Network and has also named the app Stellar Lab [12]. The main reason for the development of this lab is the use of freelancers who usually work for research purposes on the network. This lab provides all the services available on the Stellar network to develop applications that are located in the fourth layer of the Stellar Ring. In general, this network is used for the following purposes:

*1) Create test accounts with test inventory to perform the transaction*
*2) Development of applications and research and training on Stellar without the possibility of destroying valuable assets.*
*3) Testing existing applications*
*4) Perform data analysis on a smaller non-significant dataset than the original network*

In this research, we will also use this network to provide a transactional Blockchain network.

B. *Off-Chain Selection Improvement Management*

*System of Collective Distributed Destination*

Due to the limitations and protocols available in the Blockchain network, all side processes cannot be implemented on this network. Also, many of these side features by nature do not need development and cost to implement on the Blockchain. These sections will be developed as an off-chain process and will provide complementary network services to complement the functionality of the UIs. In this research, two very important parts have to be designed, developed, and integrated into the interface, which here is a wallet based on the Stellar Test Network. These sections will be developed as a web-based system outside the Blockchain network and will exchange data with the user interface using the RESTFUL API interface on the HTTP platform. These two parts are the Recognition and Customer Information Registration and the Collective Improvement Management Subsystem, based on a multi-dimensional recommender system utilizing system dimensions and users to provide targeted purposes for use by users to participate in it is voting.

**Customer recognition and registration module**
Since, in any non-public financial system, the first principle is to identify and record customer information, this service should be developed as the Blockchain network supplement to collect identity data in the first step. This subsystem generally has three main tasks, which are:

*1) User Identity Registration: Includes customer identity registration, including name, mobile number, email, national code, and public key account one by one mapping ID.*
*2) Simple mapping and logging of the public account key: The service will embed the key that was received from the Blockchain network to the off-chain network along with the unique identifier after the user logging in and making a public key for her/him.*
*3) Recording user areas of interest: This service is used to collect and record data about areas of interest for use in the selection improvement system. In general, there are four areas of interest to users of this system. These four areas include charity, education, economics, and healthcare. Users have to choose from 1 to 5 at the beginning of the network to help them with any of these areas. This selection will be used in the future as one of the targeted recommendations in the selection improvement system.*

**Selection Improvement System Module**
This is the most important part of the system that receives the information collected from the previous module as input. Then, using other dimensions implemented within the subsystem, it provides users with a targeted recommendation list with the public key of these accounts for use in the collective voting process. All the logic and algorithms involved in generating the recommendation list are implemented here. The components of this subsystem are described separately below.

A text-based recommendation system is used to generate the recommendation list for the destination selection. This model of recommenders, in addition to the basic recommendation algorithms that are based on similarity or popularity, the system incorporates factors of network and user state in their recommendations, and these factors are dynamically changeable on the network. In the system developed in this research, three important components involved in the network were considered, from each of which dimensions were selected to participate in the recommendation algorithm.

1- The first part is the purpose of being included as selected candidates in the voting process. These purposes are the entities or companies that operate in each of the four selected system areas mentioned earlier. The nature of these entities will introduce three main dimensions that will influence future decisions. These dimensions are shown in figure 4.

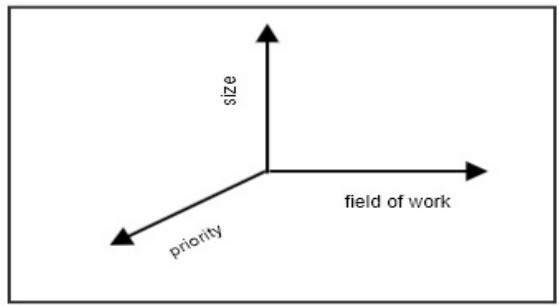

**Figure 4.** Dimensions of the system for destination entities

*a)* *The first dimension is the functional field of the entities that will be effective on users proportionate to their interests in producing voting results.*

*b)* *The second dimension is the size of the entity. The size is a number that some important factors are involved in its calculating. The number of the size in this study is determined by the size of the entity in terms of the number of served people. This number means a measure of the amount of aid needed to be absorbed to meet the needs of the entity. For example, a charity with 100 needy members has a larger size than an entity with 10 needy members. Many factors can be included in the calculation of this number, which in this research, we will use a simplified model of this factor, according to the scope of work.*

*c)* *The third dimension in this section is the system selection priority assigned by the central system management entity to each of the entities in the special condition. This factor shows, in principle, the importance of each destination entity in attracting help. The higher the number, the higher the priority and the importance of the destination entity to attract more help, and the more help should direct toward that entity. The purpose of using this factor is due to the existence of variable conditions in the real world that can make internal computing difficult. For example, if an unexpected event like an earthquake occurs, donations from people should be given to charities rather than educational entities. This factor can enhance the importance of an entity at a specific time for purposeful bidding and can be used as a lever to control internal network policy in critical times.*

2- The second part is the users in the network, at the beginning of each of them entering the network, their interest in the four areas of the system is received and recorded. This amount of interest, alongside each entity's workflow, is used as a computational dimension to generate recommendation results. Each person will be able to choose from 1 to 5 at the beginning of registration and according to their interest in each field.

3- The third part is the use of past collective selection transactions by users, which can be used as a third factor in producing targeted recommendations. This section includes a service to collect information on users' collective selection transactions. This service is used in this research to analyze the similarity of users as another factor of the system.

These three parts form the essence of the system, and targeted recommendations will be produced using these described sections and dimensions.

**The algorithm used for selection improvement section**
The algorithm consists of two general parts that together form the overall function of the system. The system receives the unique identifier described in the Customer recognition, and registration module for identifying and recording customer information at the input and the output returns a sorted list of recommendations in the JSON form that fits the user's information and attributes.

- **Basic Recommendation Generation Section:** In this section, the basic recommendation calculation algorithm is implemented. In the first step, a recommendation list is generated using the history of previous transactions of the target user and his/her similarity to other people based on the type of registrant transactions using Pearson's similarity method. The list includes several choices classified by the network's collective similarity factor. This list is then given as the input to the text-based recommendation section to improve its results based on the text parameters and to produce the final list.

- **Final Recommendation Generation:** The user's favorite list of privileges is extracted from the database using his/her unique identifier and given to the final calculation function. This function will calculate the score of this section for each entity based on a weighted average of two priority and size components as well as adding user interest based on the entity context. The calculation formula is presented in formula 3.

$$\text{Score\_CAR} = ((0.4 * \text{destinations. Priority}) + (0.6 * \text{destinations. Size}))/2 + \text{Favorite\_Category\_Score}$$

**Formula 3.** The formula for calculating the final Recommendation production section

Due to the more importance of the size factor in this study, a higher weight factor was used in the weighted average formula for this factor. Finally, the results obtained from this section are computed and combined with the results from the previous section that was obtained from the similarity of choice, and the overall score for each selected destination is calculated based on the specific characteristics of each user and returned to a user's wallet in a sorted list for final display.

### C. Wallet, base on the Stellar Network

In this study, due to the importance of gathering information from users and the necessity of having an interface through which users can interact with the network, developing a platform for users is necessary. This platform, in addition to connecting to the Stellar Blockchain network
and building an account for users on the network that includes their key pair, it should be able to connect to the off-chain sections which developed for selection improving the system and use its existing services. In the end, this interface should allow users to view targeted suggestions made in the system and use them in the cumulative voting process. To do this, we will develop a mobile application under the Stellar platform in this research. This program will be able to communicate with the test network using the Stellar libraries and perform Blockchain operations. This open-source library has been developed by Stellar for use by developers.

### Wallet Design Challenges

In any customer service provider network, an interface must be available to enable users to use the services which are embedded for them. This interface can be ATMs, store terminals, or even bank branches that are a physical interface. In Blockchain-based systems, this interface is the wallet that provides users with all the services in the financial system, including Blockchain network services or off-chain lateral services. Users of the system only see this part, and this wallet is the gate of users' information collecting. As a result, this wallet should primarily be able to communicate with the Blockchain network and provide users with basic transactions and services. For this, all Blockchain network services must be implemented based on their protocols in the wallet. In this system, in the first step, all network communication interfaces must be implemented to register and generate a pair of account keys and register encrypted transactions based on Stellar SDK. Next, all the services needed to collect data from users and provide lateral services that are not implemented by default on the Stellar network should be separately implemented and put in the wallet. After connecting both parts of the interface, the outward interface design of the program must be done so that the system can be used by users.

### Wallet Components

- **User Account Creation and Users' Basic Information Collection:** Any user who wants to log in for the first time and use the network and its services must first register on the network using her/his application and select a unique ID for her/himself. In this section, the user identity data and the ID selected by the user is received. This ID must be unique and not used by another user in the past, so this ID will be reviewed before registering information, and an error will be displayed when duplication occurs. Next, the users should record their interest in each of the fields in the system, and by using the algorithms described in the previous sections, the recommendation score is calculated. In this section, users can assign a score of 1 to 5 to each of these domains already defined in the system based on their interest. Figure 5 shows the KYC and identity data entry section of the wallet application in the android platform.

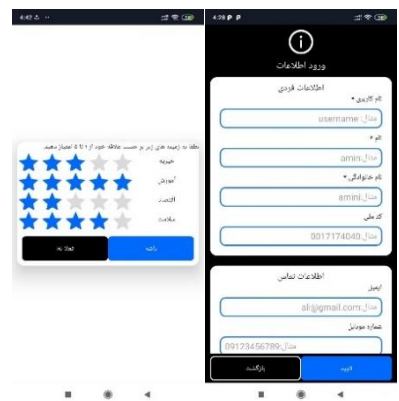

**Figure 5.** Wallet Identity Data Entry Section

- **Wallet Operations Selection:** At this step, the user enters the environment of the main wallet and will be able to perform network-based operations as well as use a cumulative voting system. This section serves as the wallet's homepage, and the user will be able to select the actions he/she wants to do after opening the wallet. These include sending or receiving money, reviewing transaction history, or participating in a collective voting process. Figure 6 shows this section of application.

## IV. FINDINGS

In this section, we try to put all the system parts together and test the functionality and effectiveness of our destination selection improvement system. In these experiments, an Android smartphone is used as the host of an Android wallet application that is the same user interface. The off-chain server also is run locally on a laptop using the Python web server. The laptop and smartphone are plugged into a local area network, and the IP address of the server is inserted into the wallet to connect the wallet to the off-chain server. The Stellar network interface server address is then placed in the wallet to use the Blockchain network services. After this process, the systems are ready to run, and the experiments can be performed on them. The specifications of the systems used in this experiment are shown in table 2.

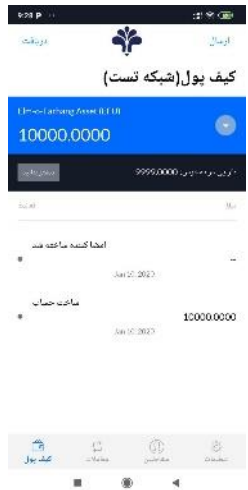

**Figure 6.** The main part of selecting a network operation in the wallet

- **Target Selection Section for Collective Selection:** This section includes the selection interface for selecting a division collective destination. In this section, users will be able to vote by declaring a public key account that they tend the collective asset belong to it. Besides, the interface to the destination selection improvement system is provided here. Upon logging into the recommendation section, users will be able to view the list of recommendations produced using their attributes. In this section, the web service of generating recommendation is recalled by sending the user ID as input, and the result of this section will show in a selectable list where the user will be able to select these destinations in a collective vote. Figure 7 shows the destination selection part of application. The left image shows the list of recommendations sorted by the scores of them and the right image is the user interface of input the selected destination of voting process.

-

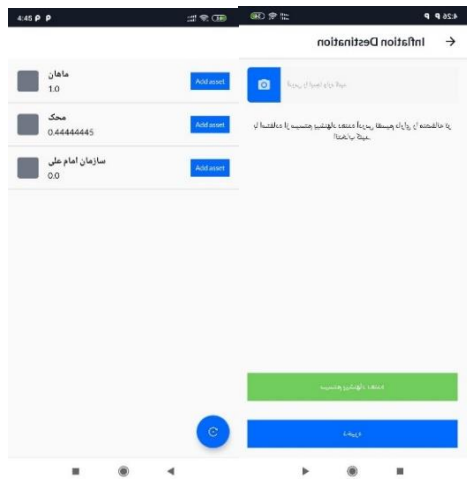

**Figure 7.** Target selection section of the collective selection destination

*1) System assumptions*

| Objective section | Executor device | Platform Key Features |
|---|---|---|
| Stellar Blockchain network | Stellar test network | 3 Servers forming the network |
| Off-chain network | Local network on MacBook Pro retina 13 | − Local router<br>− Mac operating system<br>− Python webserver |
| Mobile wallet | Android smartphone Xiaomi F1 | Android 10 |

In this section, the information recorded in the system that will be used in the upcoming testing processes is described. This information is recorded in the server as a user on the network using the wallet interface in accordance with the procedure described in the previous section. The information is shown in table 3, table 4 and table 5:

*1) Field of Activity of Selected Destination Entities in the System:*

**Table 2.** system Information used in the experiment

**Table 3.** Fields of activity of selected destination entities in the system

| Field | Context field id |
|---|---|
| Charity | 1 |
| Education and Training | 2 |
| Economic | 3 |
| Healthcare | 4 |

*2) The information about the users who have registered in the system is assumed as follows:*

*3) System Participant Purposes on the System as existing Destinations on the Improvement Selection System:*

**Table 4.** Assumed Registered Users of the System

| ID | Name | Unique ID | PublicKey |
|---|---|---|---|
| 1 | Test-one | User-1 | ANDOISQKX |
| 2 | Test-two | User-2 | KODOISQF7 |
| 3 | Test-three | User-3 | LPDOISQF7 |
| 4 | Test-four | User-4 | RTDOISQF7 |

**Table 5.** Assumed Registered Users of the System

| Entity Name | Entity Public Key | Field of Activity |
|---|---|---|
| Charity Organization | GAIRWQ | 1 |
| Education Organization | CAIRWQ | 4 |
| Economic Organization | BAIRWQ | 2 |
| Healthcare Organization | NNIRWQ | 3 |

Using these assumed data recorded in the system, we will study the following experiments and their results.

### D. Experiments in terms of size and priority of the organization

In this experiment, without changing other factors in the network for users, we will change the variables related to the size and priority of the organization and examine its impact on the results. In this experiment, the level of user interest in all fields is considered equally and unchanged, and the similarity in the network is ignored. The experiment will be conducted in three steps with different data. The sample data used in these steps are shown in table 6, table 7 and table 8.

**Table 6.** Sample Data No. 1

| name | walletID | category | priority | size |
|---|---|---|---|---|
| charity | GAIRWQ | 1 | 1 | 6 |
| health | CAIRWQ | 4 | 2 | 5 |
| education | BAIRWQ | 2 | 2 | 4 |
| economy | NNIRWQ | 3 | 2 | 1 |

**Table 7.** Sample Data No. 2

| name | walletID | category | priority | size |
|---|---|---|---|---|
| charity | GAIRWQ | 1 | 1 | 6 |
| health | CAIRWQ | 4 | 6 | 5 |
| education | BAIRWQ | 2 | 2 | 4 |
| economy | NNIRWQ | 3 | 2 | 1 |

**Table 8.** Sample Data No. 3

| name | walletID | category | priority | size |
|---|---|---|---|---|
| charity | GAIRWQ | 1 | 1 | 6 |
| health | CAIRWQ | 4 | 6 | 5 |
| education | BAIRWQ | 2 | 2 | 4 |
| economy | NNIRWQ | 3 | 2 | 13 |

The results of scoring and prioritizing the available destinations using this data are reflected in table 9.

**Table 9.** Score Results and Prioritization of Destinations in Experiment 1

|  | Test 1 | Test 2 | Test 3 |
|---|---|---|---|
| Charity Organization | 1 | 0.65 | 0.14 |
| Education Organization | 0.92 | 0.45 | 0 |
| Economic Organization | 0.69 | 0 | 1 |
| Healthcare Organization | 0 | 1 | 0.4 |

The diagram of this experiment will be as figure 8.

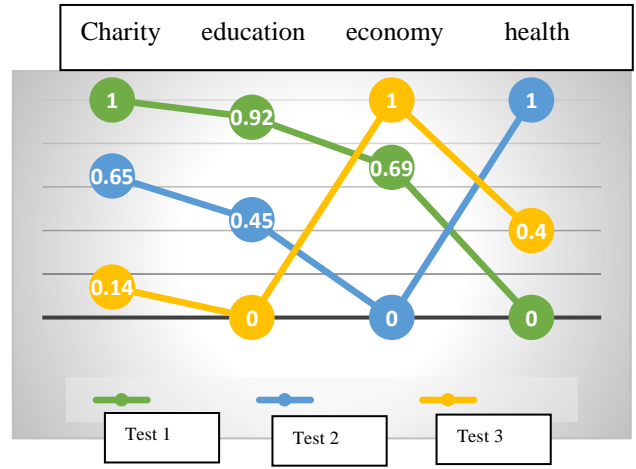

**Figure 8.** Diagram of test results by size and priority dimensions

Based on the results and the diagram, it is shown that in the developed system, increasing the size of the entity will have a direct impact on the ranking of the choices, but the effect is less than the change in priority. These scores are normally obtained from all values, and the first candidate in this section will have a score of 1, and the last candidate will have a score of 0.

### E. Experiments in terms of interest in the field of work

In this experiment, without changing other factors in the network for users, we change the variables related to the level of user interest in the organization's work fields and examine the extent to which it affects the results. The experiment is conducted in three steps with different data which, the results are shown in table 10, table 11 and table 12.

**Table 10.** Sample Data No. 1 Experiment 2

| federationID | charity | education | economy | health |
|---|---|---|---|---|
| user-1 | 1 | 1 | 3 | 4 |

**Table 11.** Sample Data No. 2 Experiment 2

| federationID | charity | education | economy | health |
|---|---|---|---|---|
| user-1 | 1 | 5 | 2 | 4 |

**Table 12.** Sample Data No. 3 Experiment 2

| federationID | charity | education | economy | health |
|---|---|---|---|---|
| user-1 | 3 | 2 | 1 | 2 |

The results of scoring and prioritizing the available destinations using this data are presented in table 13.

**Table 13.** Score Results and Destination Prioritization in Experiment 2

|  | Test 1 | Test 2 | Test 3 |
|---|---|---|---|
| Charity Organization | 0 | 0 | 1 |
| Education Organization | 0 | 1 | 0.5 |
| Economic Organization | 0.66 | 0.25 | 0 |
| Healthcare Organization | 1 | 0.75 | 0.5 |

The graph of this experiment will be as figure 9.

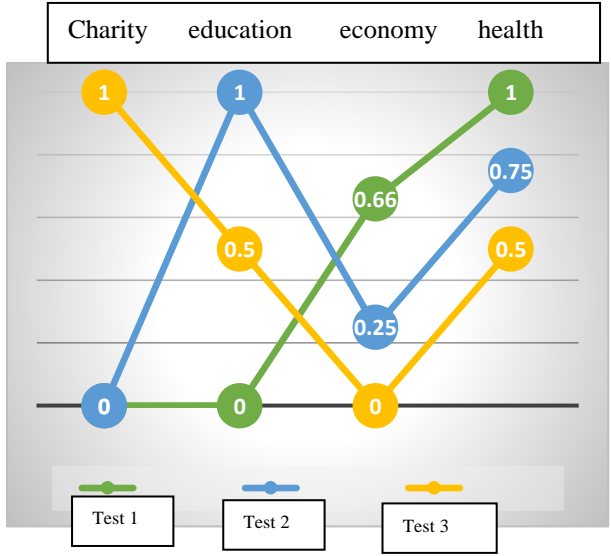

**Figure 9.** Graph of test results of interest dimension

Based on the results and the diagram, we find that in the developed selection system, the user's interest in the work of the entities in the system increases their scoring directly in the voting process. These scores are normally obtained from all values , and the first candidate in this section will have a score of 1, and the last candidate will have a score of 0.

*F. Testing the similarity of users within the network*

In this section, we examine the impact of users' similarity in the collaborative sector on the improvement recommender system. For this experiment, two data models are used, which are the data of hypothetical user-1 and user-2. The data is a list of past user transactions to participate in the voting process. Table 14 and table 15 show these sample data.

**Table 14.** Sample Data No. 1 Experiment 3

| destinationID | federationID ▲ 1 |
|---|---|
| GAIRWQ | user-1 |
| GAIRWQ | user-2 |
| BAIRWQ | user-2 |
| GAIRWQ | user-2 |
| BAIRWQ | user-2 |
| BAIRWQ | user-2 |
| CAIRWQ | user-2 |
| GAIRWQ | user-2 |

**Table 15.** Sample Data No. 2 Experiment 3

| destinationID | federationID ▼ 1 |
|---|---|
| BAIRWQ | user-2 |
| GAIRWQ | user-2 |
| BAIRWQ | user-2 |
| CAIRWQ | user-2 |
| CAIRWQ | user-2 |
| GAIRWQ | user-1 |

The results of scoring and prioritizing the available destinations using this data are shown in table 16.

**Table 16.** Score Results and Prioritization of Destinations in Experiment 3

|  | Test 1 | Test 2 |
|---|---|---|
| Charity Organization | 0 | 0 |
| Education Organization | 1 | 1 |
| Economic Organization | 0 | 0 |
| Healthcare Organization | 0 | 1 |

The results show that in this section, the similarity between system users in the number and nature of the destinations they have chosen will be directly able as a dimension to influence the scores and priorities, and this will allow the users to get targeted results in their recommendations. These results make it easier for users to participate in the process and to guide their choices.

These results show that the users' choices are primarily dependent on the likelihood of them with other users of the network based on their preferences, and these results are then influenced by other dimensions of the system and determine the overall decision direction. Based on the results of the experiments carried out and the results obtained, it can be concluded that the selection of a context-aware recommender system that can act as a complementary subsystem to the Stellar financial networks can serve as a useful solution to manage and improve the collective voting process in these networks. Using the dimensions of the network and users, the system can improve users' choices and manage them to make fairer choices. Since the targeted division of collective assets is critical in these networks, this approach can greatly help address this challenge.

## V. CONCLUSION

One of the most important challenges in the Stellar network is to improve the distribution of collective assets based on voting in the network. This is very important concerning the high volume of funds collected and distributed in this sector. In this research, after identifying a network and launching an example of it, we attempted to identify a network based on a text-based recommender system to solve one of the challenges that exist, collective asset division. To do this, first, a system with a standard structure had to be created using the critical components of the system. In this research, we developed a mobile wallet on the Android platform using the libraries provided by Stellar Network. This wallet will be able to connect to off-chain server services as well as providing complementary services, which is here a collective improvement system, in addition to connecting to Stellar's web services. In this research, in addition to designing and implementing a complete financial Blockchain network consisting of all the necessary elements, we present a solution using recommender systems to solve one of the challenges in this network, which is the purposeful collective asset division. This system helps users to participate fairer in the voting process by making targeted suggestions using the dimensions of the system. Then, a number of tests were reviewed on the system to demonstrate the effectiveness of this method for solving the existing challenges. The results of the experiments show that the use of recommender systems is an appropriate solution to the challenge of the targeted division of collective assets in the Stellar. Concerning the importance of this issue in real-world financial networks and the growing development of Blockchain technology among payment research firms in future research, the following issues can be addressed in the context of this project:

*1)* *Research on the network transaction analysis to provide an algorithm to more accurately calculate the dimensions involved in the system.*

*2)* *Involve other dimensions of the network to improve the recommendation process.*

**REFERENCES**

[1] Guo, Y., & Liang, C. (2016). Blockchain application and outlook in the banking industry. Financial Innovation, 2(1), 24.
[2] Feng, Q., He, D., Zeadally, S., Khan, M. K., & Kumar, N. (2019). A survey on privacy protection in blockchain system. Journal of Network and Computer Applications, 126, 45-58.
[3] Zheng, Z., Xie, S., Dai, H., Chen, X., & Wang, H. (2017, June). An overview of blockchain technology: Architecture, consensus, and future trends. In 2017 IEEE international congress on big data (BigData congress) (pp. 557-564).
[4] Mazieres, D. (2015). The stellar consensus protocol: A federated model for internet-level consensus. Stellar Development Foundation, 32.
[5] Stellar Team. (2018, October 6). Stellar Inflation. Retrieved from https://www.stellar.org/developers/guides/concepts/inflation.html
[6] Gajane, P., & Pechenizkiy, M. (2017). On formalizing fairness in prediction with machine learning. arXiv preprint arXiv:1710.03184.
[7] Zafar, M. B., Valera, I., Rodriguez, M. G., & Gummadi, K. P. (2015). Fairness constraints: Mechanisms for fair classification. arXiv preprint arXiv:1507.05259.
[8] Hardt, M., Price, E., & Srebro, N. (2016). Equality of opportunity in supervised learning. In Advances in neural information processing systems (pp. 3315-3323).
[9] Kim, M., Reingold, O., & Rothblum, G. (2018). Fairness through computationally-bounded awareness. In Advances in Neural Information Processing Systems (pp. 4842-4852).
[10] Rajput, I. S., & Gupta, D. (2012). A priority-based round robin CPU scheduling algorithm for real-time systems. International Journal of Innovations in Engineering and Technology, 1(3), 1-11.
[11] Zehlike, M., Bonchi, F., Castillo, C., Hajian, S., Megahed, M., & Baeza-Yates, R. (2017, November). Fa* ir: A fair top-k ranking algorithm. In Proceedings of the 2017 ACM on Conference on Information and Knowledge Management (pp. 1569-1578).
[12] Kim, M., Kwon, Y., & Kim, Y. (2019, June). Is Stellar as secure as you think?. In 2019 IEEE European Symposium on Security and Privacy Workshops (EuroS&PW) (pp. 377-385).

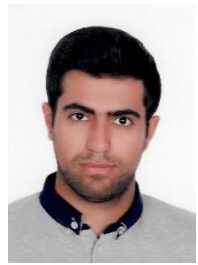

**Kiarash Shamsi** has received the master of software engineering from the department of computer engineering at the University of Science and Culture, Tehran, Iran. His research interests include Blockchain, data science and distributed systems.

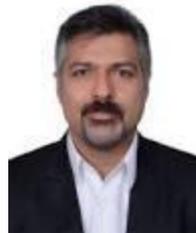

**Mohammad Javad Shayegan** is an Associate Professor at the Department of Computer Engineering at the University of Science and Culture, Tehran, Iran. He is the founder of the web research center, International Conference on Web Research, and International Journal of Web Research in Iran. His research interests include Web research, data science and distributed systems